# DeepResp: Deep learning solution for respiration-induced $B_0$ fluctuation artifacts in multi-slice GRE

Hongjun An[1], Hyeong-Geol Shin[1], Sooyoen Ji[1], Woojin Jung[1], Sehong Oh[2], Dongmyung Shin[1], Juhyung Park[1], and Jongho Lee[1]

**Author affiliations:**

[1]Laboratory for Imaging Science and Technology, Department of Electrical and Computer Engineering, Seoul National University, Seoul, Korea

[2]Division of Biomedical Engineering, Hankuk University of Foreign Studies, Gyeonggi-do, Korea

**Corresponding Author:**

Jongho Lee, Ph.D

Department of Electrical and Computer Engineering, Seoul National University

Building 301, Room 1008, 1 Gwanak-ro, Gwanak-gu, Seoul, Korea

Tel: 82-2-880-7310

E-mail: jonghoyi@snu.ac.kr

**ABSTRACT**

Respiration-induced $B_0$ fluctuation corrupts MRI images by inducing phase errors in k-space. A few approaches such as navigator have been proposed to correct for the artifacts at the expense of sequence modification. In this study, a new deep learning method, which is referred to as DeepResp, is proposed for reducing the respiration-artifacts in multi-slice gradient echo (GRE) images. DeepResp is designed to extract the respiration-induced phase errors from a complex image using deep neural networks. Then, the network-generated phase errors are applied to the k-space data, creating an artifact-corrected image. For network training, the computer-simulated images were generated using artifact-free images and respiration data. When evaluated, both simulated images and *in-vivo* images of two different breathing conditions (deep breathing and natural breathing) show improvements (simulation: normalized root-mean-square error (NRMSE) from 7.8 ± 5.2% to 1.3 ± 0.6%; structural similarity (SSIM) from 0.88 ± 0.08 to 0.99 ± 0.01; ghost-to-signal-ratio (GSR) from 7.9 ± 7.2% to 0.6 ± 0.6%; deep breathing: NRMSE from 13.9 ± 4.6% to 5.8 ± 1.4%; SSIM from 0.86 ± 0.03 to 0.95 ± 0.01; GSR 20.2 ± 10.2% to 5.7 ± 2.3%; natural breathing: NRMSE from 5.2 ± 3.3% to 4.0 ± 2.5%; SSIM from 0.94 ± 0.04 to 0.97 ± 0.02; GSR 5.7 ± 5.0% to 2.8 ± 1.1%). Our approach does not require any modification of the sequence or additional hardware, and may therefore find useful applications. Furthermore, the deep neural networks extract respiration-induced phase errors, which is more interpretable and reliable than results of end-to-end trained networks.

**Introduction**

In MRI, respiration is known as a source for physiology-induced noises that introduce not only physical motion but also $B_0$ field fluctuation (Noll and Schneider, 1994; Raj et al., 2000; Versluis et al., 2010; Wen et al., 2015). This fluctuation is induced by the chest motion during respiration, producing approximately 0.01 ppm change in the brain (e.g. 1.3 Hz at 3 T) (Versluis et al., 2010; Wowk et al., 1997). The change generates phase errors in the k-space of gradient echo (GRE) images, mostly along the phase encoding direction, creating image artifacts or ghosting along the direction. The fluctuation was reported to be homogeneous across an axial slice and decrease with the distance from the lungs (Van de Moortele et al., 2002).

To correct for the respiration-induced artifacts, a few methods have been proposed such as using a navigator echo (Durand et al., 2001; Ehman and Felmlee, 1989; Lee et al., 2006; Wowk et al., 1997) and external tracking device (Boer et al., 2012; Duerst et al., 2015; Ehman et al., 1984; Van Gelderen et al., 2007; Vannesjo et al., 2015). These methods, however, require modification of software or hardware, limiting applications of the methods. More recently, Meineke et al. (Meineke and Nielsen, 2019) proposed another method, which only utilized coil sensitivity profiles but required complex data processing.

Recently, deep learning has been widely applied in MRI. A few deep learning-based methods have been proposed to correct for physical motion-induced artifacts in the images (Armanious et al., 2019; Duffy et al., 2018; Jiang et al., 2019; Lv et al., 2018; Tamada et al., 2020). So far, however, no study has been proposed to correct for the artifacts from the respiration-induced $B_0$ fluctuation, although a study proposed to predict the fluctuation from respiration belt data (Niklas Wehkamp and Zaitsev, 2017) and another work suggested $B_0$ calibration for glutamate weighted chemical exchange saturation transfer (Li et al., 2020).

In this paper, we demonstrate a new retrospective respiration-induced artifact correction method using deep neural networks. Different from popular end-to-end deep learning methods (Kwon et al., 2017; Lee et al., 2020; Yoon et al., 2018), our network is designed to extract phase errors from respiration-induced $B_0$ fluctuation and, therefore, the result is more interpretable. The network-generated errors are utilized to compensate for the respiration-induced phase errors in k-space, producing an artifact-corrected image. This method is referred to as DeepResp, hereafter.

**Materials and methods**

**DeepResp**

In 2D GRE, the respiration-induced $B_0$ fluctuation can be modeled as phase errors in k-space that vary along the phase encoding (PE) direction when small changes during the readout are ignored (Noll and Schneider, 1994). Our DeepResp was designed to extract these respiration-induced phase errors from a multichannel-combined complex image using deep neural networks (Fig. 1). The network-generated phase errors were conjugated and applied to the k-space of each channel, correcting for the errors. Finally, the multichannel k-space data were reconstructed to an artifact-corrected image (Fig. 1b).

DeepResp had two stages (Fig. 1c). In the first stage, the k-space data (matrix size: 224 × 224), which were generated from the multichannel-combined complex image, were divided into 14 groups, each with 16 neighboring PE lines plus one additional PE line for differential operation (see below). Each group was then processed separately to estimate the respiration-induced phase information of the group. This grouping had benefits in network performance because the neighboring PE lines had similar signal intensities, helping the network better extract the phase information than using the whole k-space data. In each group, the rest of the k-space was zero-filled (i.e. bandpass-filtered). Then, this k-space was inverse Fourier-transformed, generating a complex bandpass-filtered image (Fig. 1c). This complex image and the original input complex image were split into real and imaginary images. For the bandpass-filtered real and imaginary images, a batch-normalization-layer was applied to adjust the input range of the network. For the original input images, they were normalized such that the magnitude had a standard deviation of 0.25. Finally, the four images were concatenated along the third (or channel) dimension, generating a 224 × 224 × 4 matrix. The matrix was input into a neural network, modified ResNet50 (He et al., 2016), generating 16 "differential values" of the phase errors as the output (see Discussion). The details of the network structure are described in Supplementary Information S1a.

The second stage of DeepResp was designed to accumulate the differential values of phase errors and to correct for error accumulation in order to produce the respiration-induced phase errors. The network utilized a 1D convolutional autoencoder structure (Masci et al., 2011), of which details are described in Supplementary Information S1b. The input of the network was 224 × 1 vector, which was the output of the first stage network. The output of the second stage network had the same size vector as the input.

To generate a 2D phase compensation matrix for the phase correction of the k-space data (Fig. 1b), the output vector of the neural network was used as the phase of a complex exponential vector. After that, the complex vector was transformed into a 224 x 224 matrix by extending the vector along the read-out direction. This matrix was conjugated and then multiplied to the k-space of each channel of the input image. These phase error-corrected multichannel k-space data were reconstructed to a channel-combined complex image by multiplying the multichannel complex images with the conjugates of coil-sensitivity (Uecker et al., 2014) and then averaging the channels. The same approach was also applied when generating the multichannel-combined complex input image.

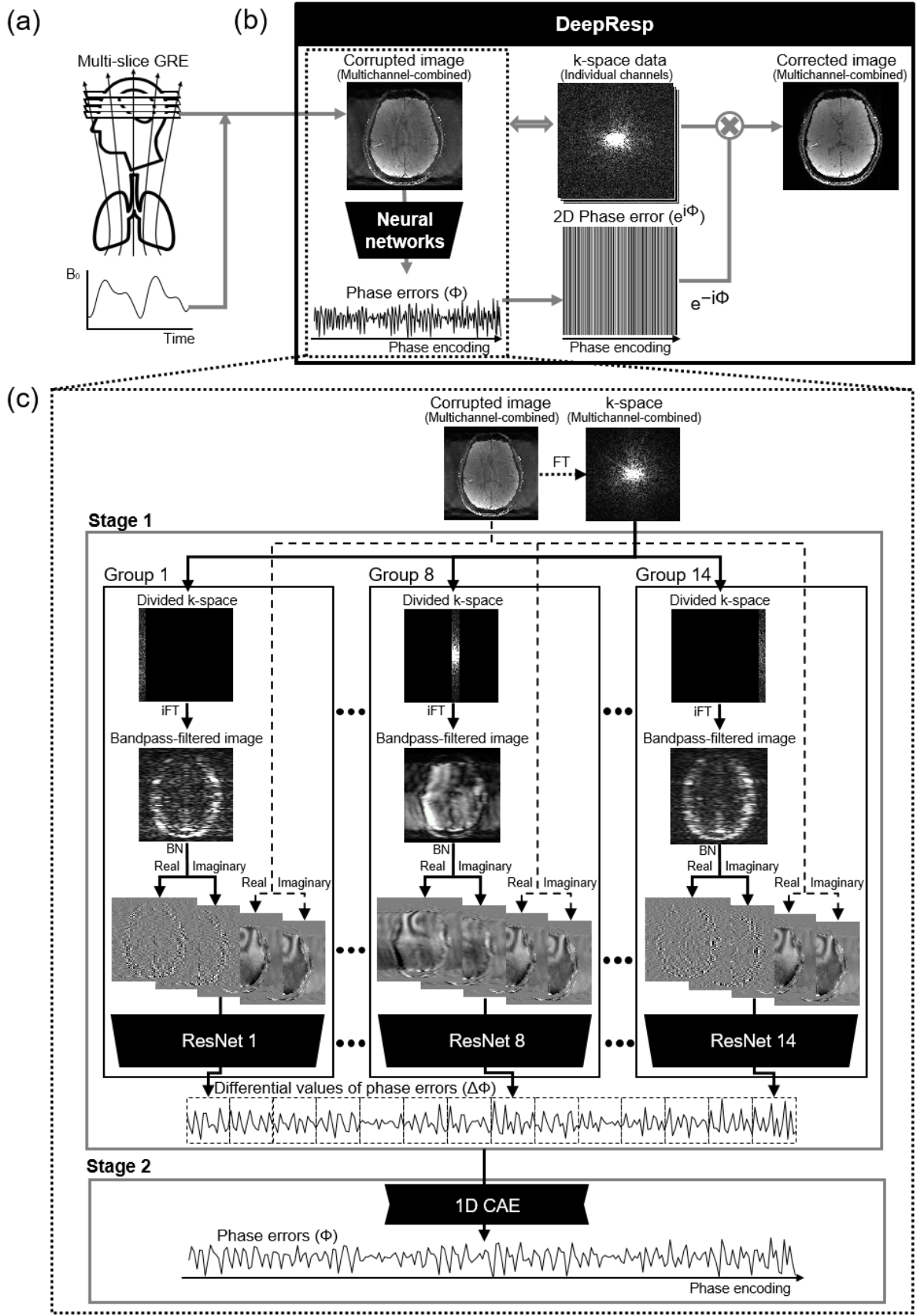

**Figure 1.** Overview of DeepResp. (a) Schematic of respiration-induced $B_0$ fluctuation in the brain. (b) Structure of DeepResp. DeepResp is designed to extract the phase error from a GRE image. (c) Architecture of a two-stage neural network in DeepResp. The first stage extracts the differential values of the phase errors. The second stage accumulates the differential values, generating the phase errors.

**Network Training**

For the training of the networks, we generated simulated respiration-corrupted complex images using existing GRE images and respiration data (Fig. 2). The GRE images were from 21 subjects (9 subjects from Yoon et al. (Yoon et al., 2018) and 12 subjects from Jung et al. (Jung et al., 2020)). The scan parameters are summarized in Supplementary Information S2. The images were either zero-padded or cropped in k-space to match the matrix size to 224 × 224. After discarding low-intensity images, a total of 1,936 complex images (2D) were utilized. Each image was masked out noises in the background using an intensity threshold to remove artifacts in the background. The respiration data from 111 subjects (data from an unpublished work) were measured using a temperature sensor (Biopac, MP150WS, Goleta, USA). The data were sampled at 500 Hz and recorded for 7 sessions, each with 390 seconds. A median filter and a bandpass-filter (passband: 0.1 Hz ~ 1 Hz) were applied to reduce noise. Out of the 1,936 images and 111-subject respiration data, 1,655 images from 18 subjects and 100-subject respiration data were used for the training, while the remaining data were used for the network evaluation.

The procedure to generate a respiration-corrupted image is as follows. First, the respiration data from a randomly-chosen subject were sampled with a sampling period of 1.2 sec for 224 points, starting at a random time point. The sampled data were scaled to have a peak amplitude of value between 0.03 rad and 0.63 rad. The latter corresponds to the frequency shift of 2.5 Hz at TE of 40 ms. The data were reformatted to a 2D phase error matrix as described in the previous section. Then, the matrix was multiplied to the k-space of a randomly-chosen 2D complex image, which was rotated by a random angle between -10° and 10°, and optionally left-right flipped before the multiplication. This respiration-corrupted k-space was inverse Fourier-transformed to generate real and imaginary images. Then, Gaussian noise was added to them, setting the signal-to-noise-ratio (SNR) of the magnitude image to be between 30 and 200. Using this procedure, a total of 1 million pairs of respiration-corrupted images and phase errors were simulated for the training and 20,000 pairs for the evaluation. The images were normalized such that the magnitude had a standard deviation of 0.25 as mentioned before. The phase errors were re-scaled to set the maximum phase error (i.e. 0.63 rad) to be 1.

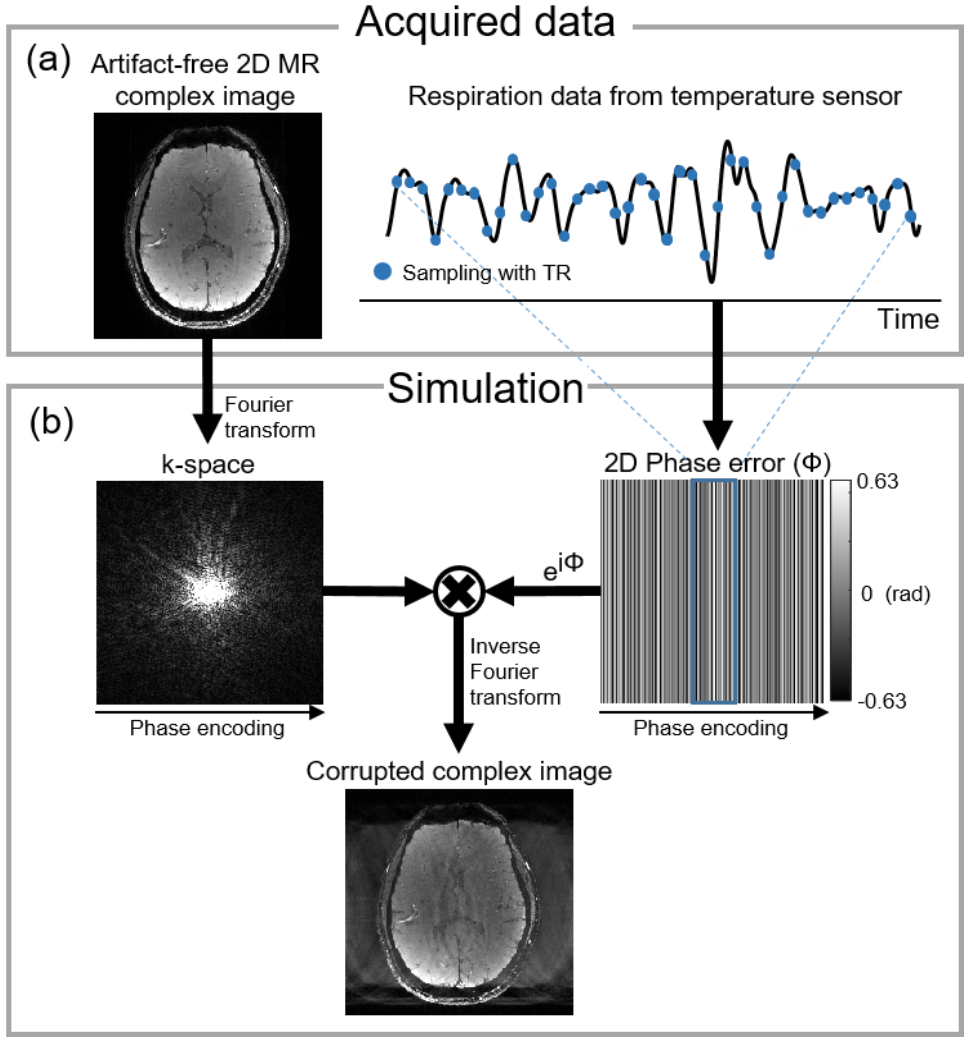

**Figure 2.** Generation of the training data using computer simulation. (a) Example of an MRI image and a respiration pattern for the computer simulation. (b) Respiration-corrupted images created by the phase modulation of the respiration pattern.

When training DeepResp, each stage was trained separately to avoid GPU memory overflow. The first stage was trained for 4 epochs and the second stage for 2 epochs. The networks were trained using ADAM optimizers (Kingma and Ba, 2014) with a learning rate scheduler (ReduceLROnPlateau; initial learning rate: $4e^{-3}$, decay factor: 0.5, patience: 1,000 iterations, and threshold: $1e^{-3}$), mean squared error loss function, batch size of 100, and initialization suggested in He et al. (He et al., 2015). Python and Pytorch library (Paszke et al., 2019) were used for programming and four Nvidia TITAN Xp GPUs (Nvidia Crop., Santa Clara, CA) for hardware.

**Network Evaluation**

The evaluation of DeepResp was performed using the simulated data and newly acquired *in-vivo* data. For the simulated data, 20,000 images reserved for the evaluation were used. For the i*n-vivo* data, 10 subjects were scanned at a 3T scanner (Trio, Siemens, Erlangen, Germany) using a multi-slice GRE sequence (Nam et al., 2015). The study was approved by the institutional review board. The subjects were instructed to breathe naturally for the first scan and then to breathe deeply for the second scan to test the two different breathing conditions. The sequence contained a navigator echo, which was used to generate reference phase errors. The scan parameters were as follows: TR = 1200 ms, TE = 6.9 ms, 15.2 ms, 20.5 ms, 25.7 ms, 31.0 ms, 36.3 ms, and 41.5 ms for the images, 55.0 ms for the navigator, flip angle = 70°, bandwidth = 260 Hz/pixel, FOV = 224 × 224 mm$^2$, in-plane resolution = 1 × 1 mm$^2$, slice thickness = 2 mm, distance factor = 20%, and 18 slices for 9 subjects and 16 slices for 1 subject. Only the last echo images, which showed the most severe respiration-induced artifacts, were used. The image reconstruction was performed offline using multichannel k-space data as described in the DeepResp section.

The navigator echo data were used to obtain the phase errors. The i$^{th}$ phase error ($\Delta\Phi_i$) was estimated by subtracting the phase of the central PE line ($\Phi_{j, 113}$ where j indicates an index for readout) from that of a PE line ($\Phi_{j, i}$) via the inner product of two complex-valued functions ($M_{j, i}M_{j, 113}e^{i(\Phi_{j, i}-\Phi_{j, 113})}$), which were summed over the read-out direction. Then, the phase of the result was averaged over the channels, generating 224 phase errors. From these phase errors, a slowly varying phase term, which may not be from the respiration, was eliminated by fitting and removing 15$^{th}$ order polynomials. The phase errors were scaled for the TE difference between the image and the navigator. Finally, the phase errors were corrected in each channel of k-space data, generating navigator-corrected images.

The respiration-induced phase errors from DeepResp were compared with the labels (simulation data) or those of the navigator (experimental data) using the Pearson correlation coefficient. Normalized root-mean-square error (NRMSE) and structural similarity (SSIM) were calculated in the uncorrected and DeepResp-corrected images with respect to a reference image, which was either the uncorrupted image in the simulation or the navigator-corrected image in the experiment. Ghost-to-signal-ratio (GSR), which was defined as the ratio of ghost intensity minus background intensity to image intensity (see Supplementary Information S3), was computed for all images (Giannelli et al., 2010).

**Results**

DeepResp successfully suppressed the respiration-induced artifacts in the simulated data (Fig. 3). The correlation coefficient of the DeepResp-estimated phase errors and the label was very high (Fig. 3d, h, l, p), reporting 0.92 ± 0.12 for all the test data. The image quality metrics were substantially improved after the DeepResp processing (NRMSE: from 7.8 ± 5.2% to 1.3 ± 0.6%; SSIM: from 0.88 ± 0.08 to 0.99 ± 0.01; GSR: from 7.9 ± 7.2% to 0.6 ± 0.6%; averages of all images).

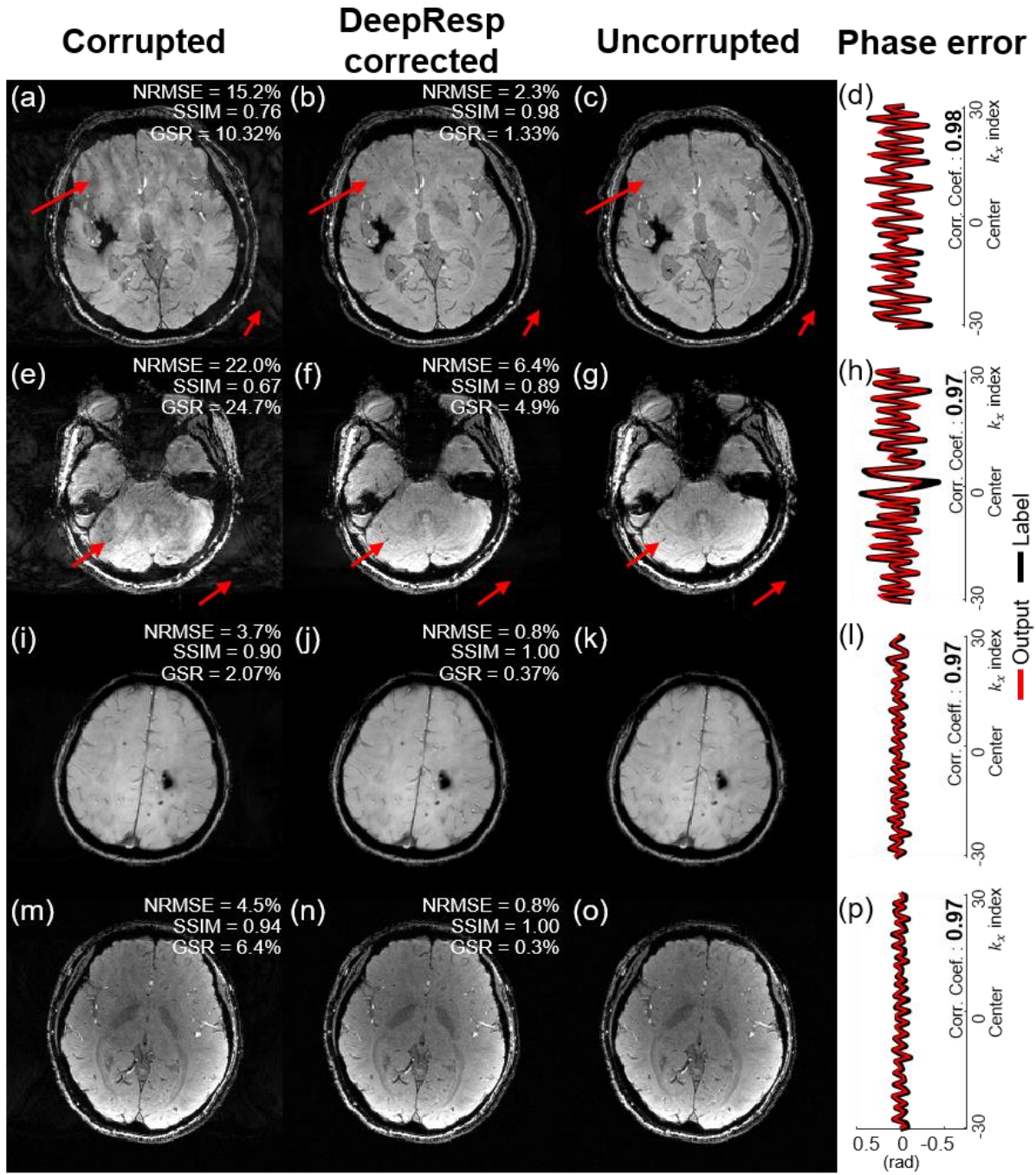

**Figure 3.** DeepResp results for the simulated data. The corrupted images (first column), the DeepResp-corrected images (second column), the uncorrupted images (third column), and the phase errors (last column) are displayed. For large phase errors (first and second rows), the artifacts are clearly observed in the corrupted images (a, e; red arrows) and Deepresp-corrected images reveal little artifacts (b, f; red arrows). For small phase errors (third and fourth rows), the artifacts are not visible but the metrics show improvements after the correction. The phase errors in (d, h, l, and p) report phase values near the center of the k-space, showing a high correlation between the label (black line) and the DeepResp result (red line).

When DeepResp was applied to the *in-vivo* data of the two different breathing conditions, the results also showed improvements (Fig. 4 for deep breathing and Fig. 5 for natural breathing). In the deep breathing condition, the uncorrected images clearly revealed artifacts that were removed in both DeepResp-corrected and navigator-corrected images (Fig. 4a, h: red arrows for the artifacts outside of the brain, Fig. 4d, k: yellow arrows for the artifacts inside of the brain). The correlation coefficient of the phase errors between the outputs of DeepResp and navigator was $0.83 \pm 0.13$ (all subjects). The mean quantitative metrics of all subjects were substantially improved after the correction using DeepResp (NRMSE: from $13.9 \pm 4.6\%$ to $5.8 \pm 1.4\%$; SSIM: from $0.86 \pm 0.03$ to $0.95 \pm 0.01$; GSR: $20.2 \pm 10.2\%$ to $5.7 \pm 2.3\%$). The mean GSR for the navigator-corrected images was $4.5 \pm 2.2\%$, which was similar to that of DeepResp ($5.7 \pm 2.3\%$). In the natural breathing condition, the respiration-induced artifacts were subtle and were difficult to identify within the brain (Fig. 5a, h). When the display range was reduced by a factor of 10, however, artifacts in the outside of the brain were visible (Fig. 5d, k). These artifacts were substantially reduced after the corrections (red arrows in Fig. 5). The average correlation coefficient of the phase errors was $0.55 \pm 0.24$ (all subjects). The mean quantitative metrics of all subjects showed improvements (NRMSE: from $5.2 \pm 3.3\%$ to $4.0 \pm 2.5\%$; SSIM: from $0.94 \pm 0.04$ to $0.97 \pm 0.02$; GSR: $5.7 \pm 5.0\%$ to $2.8 \pm 1.1\%$). The mean GSR for the navigator-corrected images was $1.5 \pm 0.9\%$. Table 1 summarizes the results of the quantitative metrics. The results of all individuals are included in Supplementary Information S4 to S8.

**Table 1.** Means NRMSE, SSIM, and GSR for all subjects. The navigator-corrected images were used as reference images.

|  |  | Uncorrected | DeepResp-corrected | Navigator-corrected |
|---|---|---|---|---|
| Deep breathing | NRMSE (%) | 13.9 ± 4.6 | 5.8 ± 1.4 | - |
|  | SSIM | 0.86 ± 0.03 | 0.95 ± 0.01 | - |
|  | GSR (%) | 20.2 ± 10.2 | 5.7 ± 2.3 | 4.5 ± 2.2 |
|  | Correlation coefficient | - | 0.83 ± 0.13 | - |
| Natural breathing | NRMSE (%) | 5.2 ± 3.3 | 4.0 ± 2.5 | - |
|  | SSIM | 0.94 ± 0.04 | 0.97 ± 0.02 | - |
|  | GSR (%) | 5.7 ± 5.0 | 2.8 ± 1.1 | 1.5 ± 0.9 |

| Correlation coefficient | - | 0.55 ± 0.24 | - |

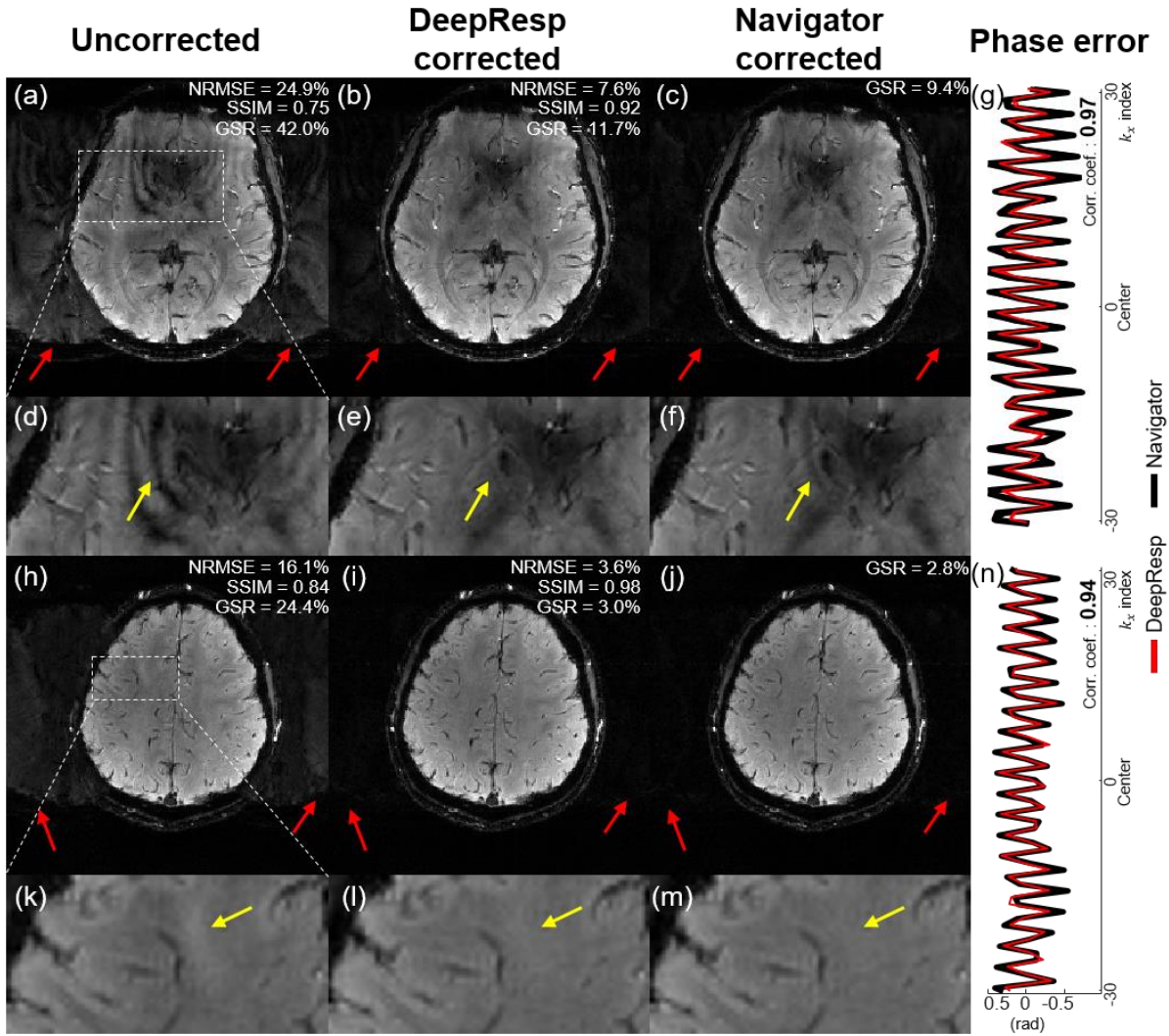

**Figure 4.** *In-vivo* results of DeepResp in deep breathing. Two slices (first and third rows) and their zoomed-in images (second and fourth rows) are shown. Artifacts (red and yellow arrows) are clearly reduced in DeepResp- and navigator-corrected images. The phase errors show very high correlations between the results of DeepResp (red line) and the navigator (black line).

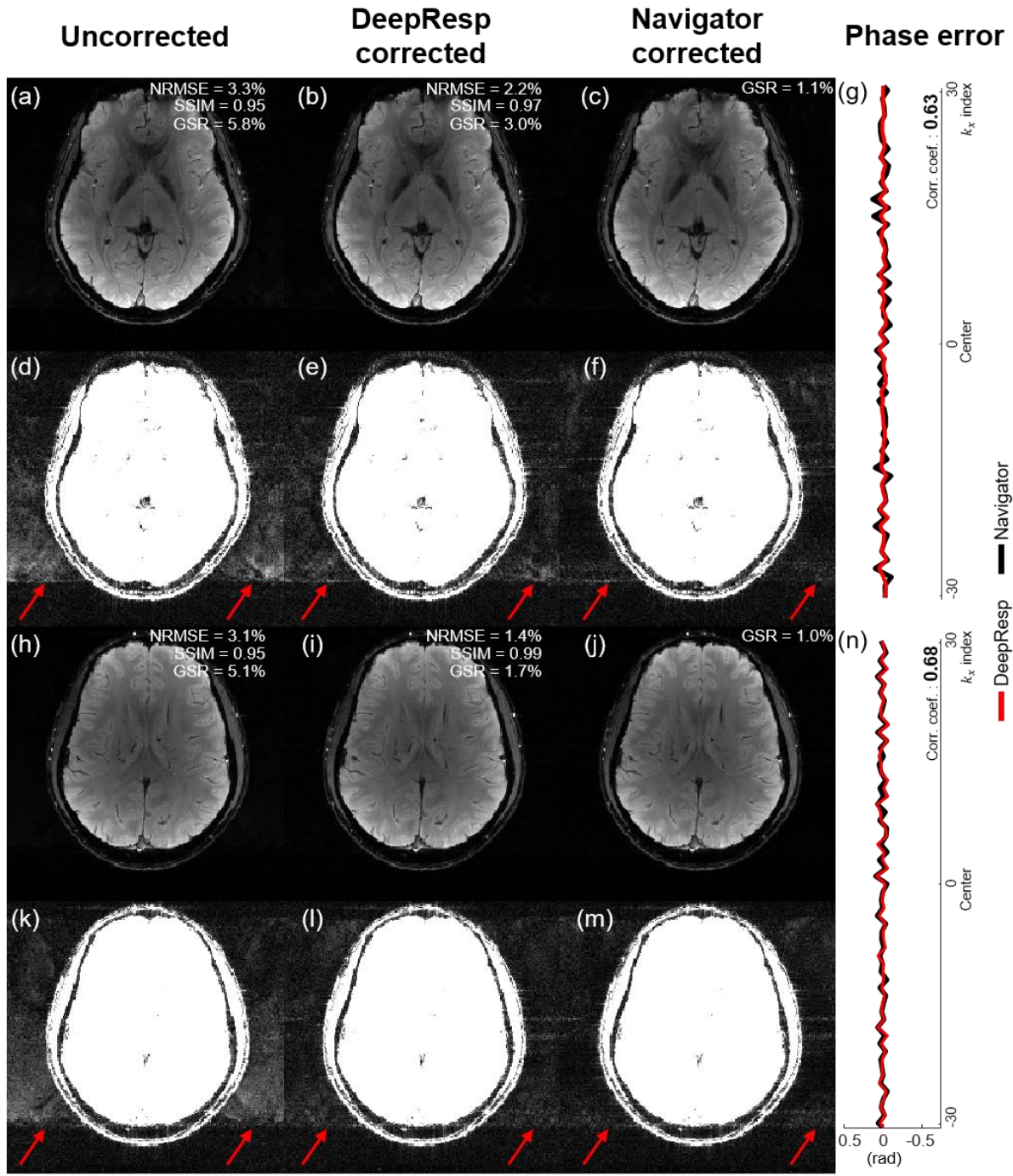

**Figure 5.** *In-vivo* results of DeepResp in natural breathing. To illustrate subtle respiration-induced artifacts, the display range was reduced by 1/10. Both slices showed decreased NRMSE, increased SSIM, and reduced GSR, suggesting successful correction of the artifacts. The phase errors from DeepResp also show high correlations with those of the navigator.

**Discussion and Conclusion**

In this work, a new deep learning-powered artifact correction method that compensated for the $B_0$ fluctuation from respiration was proposed. The method extracted the respiration-induced phase errors from a multi-slice GRE image with no additional information. These phase errors were applied in k-space to generate an artifact-corrected image. The results revealed significantly reduced respiration-induced artifacts in both simulated and *in-vivo* images. This outcome has important value in applications because DeepResp needs no external hardware or modification of the sequence. As compared to other end-to-end based deep learning methods for artifact corrections, our network is designed for a well-defined function (i.e., extraction of a respiration pattern), and the result is interpretable. This point may have great importance for the reliability of the method.

In this paper, DeepResp was evaluated with the fixed scan parameters. The method may accommodate different TEs, flip angles, resolution, etc. (see the results of different TEs in Supplementary Information S9). For a different matrix size and TR, however, modifications are necessary for the networks. For example, a larger size matrix needs an increased number of groups or PE lines per group, which will require re-training of the network.

DeepResp extracted the respiration-induced phase errors from a multichannel-combined complex image, while the correction of the phase errors required the k-space data of all channels. When the correction was applied to the channel-combined data, the results were degraded, limiting our ability to correct for the artifacts in channel-combined Dicom images. For the generation of multichannel-combined complex images, we utilized ESPIRiT (Uecker et al., 2014). Other approaches of combining multichannel data (Bernstein et al., 1994; Hammond et al., 2008; Robinson et al., 2011) resulted in different performances, probably because they generated different artifact patterns.

In the first stage of DeepResp, k-space was divided into a few groups to improve the performance. When grouping was not utilized, the information in the edges of the k-space could not be extracted because of its low intensity, producing less effective correction results. The grouping, however, dramatically increased the size of the network. As an alternative approach for grouping, one may utilize a filter that equalizes k-space intensities (Han et al., 2019).

In our work, the differential values of the phase errors were used for the training of the first stage to improve network performance. This improvement may be explained by the fact that the differential value of the phase determined the relative shift of the image. When compared with the results using the phase errors directly, those using the differential values substantially improved the network training convergence speed and network performance.

The current implementation of DeepResp is designed for multi-slice 2D GRE images. Generalization for 3D GRE images may require several modifications because the effects of respiration for 3D k-space encoding with two PE directions are different from those in 2D. Further research is necessary to find a solution for 3D.

DeepResp may find important applications in body imaging and ultra-high-field imaging (e.g. 7T) because of increased respiration-artifacts (Bolan et al., 2004; Van de Moortele et al., 2002). Applications to these areas may expand the utility of the proposed method.

**Acknowledgments**

This work was supported by the National Research Foundation of Korea (NRF) grant funded by the Korea government (MSIT) (NRF-2018R1A2B3008445) and the Brain Research Program through the NRF funded by the Ministry of Science, ICT & Future Planning (NRF-2015M3C7A1031969). The Institute of Engineering Research at Seoul National University provided research facilities for this work.